\documentclass[12pt]{article}
\usepackage{amssymb}
\usepackage{amsmath}
\usepackage{epsfig}
\usepackage{float}
\newcommand{\be}{\begin{equation}}
\newcommand{\ee}{\end{equation}}
\newcommand{\bea}{\begin{eqnarray}}
\newcommand{\eea}{\end{eqnarray}}
\begin{document}

\begin{center}
{\bf Recoilless resonant neutrino capture and basics of neutrino
oscillations}
\end{center}
\begin{center}
S. M. Bilenky
\end{center}

\begin{center}
{\em  Joint Institute for Nuclear Research, Dubna, R-141980,
Russia\\}
\end{center}
\begin{center}
F. von  Feilitzsch and W. Potzel
\end{center}
\begin{center}
{\em Physik-Department E15, Technische Universit\"at M\"unchen, D-85748
Garching, Germany}
\end{center}
\begin{abstract}
It is shown that the experiment on recoilless resonant emission and absorption of
$\bar \nu_{e}$, proposed recently by Raghavan, could have an
important impact on our understanding of the physics of neutrino
oscillations.
\end{abstract}
\section{Introduction}
Evidence for neutrino oscillations obtained in the Super-Kamiokande atmospheric \cite{SK}, SNO solar  \cite{SNO}, KamLAND reactor\cite{Kamland} and other neutrino experiments
\cite{Cl,Gallex,Sage,SKsol} is one of the most important recent
discoveries in particle physics. There is no natural explanation
of the smallness of neutrino masses and of the large mixing angles in the Standard Model (SM). There is
a general opinion that small neutrino masses and neutrino mixing are
signatures of new physics beyond the SM.

All existing neutrino oscillation data with the exception of the LSND
data \cite{LSND}\footnote{Indication in favor of
$\bar\nu_{\mu}\leftrightarrows\bar\nu_{e} $
oscillations obtained in
the accelerator short-baseline LSND experiment are going to be
checked by the running MiniBooNE experiment \cite{Miniboone}.} are
in a good agreement with three-neutrino mixing. In the framework
of three-neutrino mixing, from the analysis of the Super-Kamiokande
atmospheric neutrino data the following ranges for the largest neutrino mass-squared
difference $\Delta m^2_{23}$ and for the mixing angle were obtained
\cite{SK}\footnote{ Neutrino mass-squared difference is given by
$\Delta m^2_{ik}=m^2_{k}-m^2_{i}.$}
\begin{equation}\label{1}
 1.5\cdot 10^{-3}\leq \Delta m^{2}_{23} \leq 3.4\cdot
10^{-3}\rm{eV}^{2};~~\sin^{2}2\theta_{23}>0.92
\end{equation}
From a global analysis of the KamLAND and solar neutrino data it was found \cite{Kamland}
\begin{equation}\label{2}
\Delta m^{2}_{12} =
7.9^{+0.6}_{-0.5}~10^{-5}~\rm{eV}^{2};~~\tan^{2}\theta_{12}=0.40^{+0.10}_{-0.07}.
\end{equation}
The investigation of neutrino oscillations is based on the following
assumptions:

\begin{enumerate}
\item
Neutrino interactions are the SM charged current (CC) and neutral current (NC)
interactions. The leptonic CC and neutrino NC are given by
\begin{equation}\label{3}
j^{\mathrm{CC}}_{\alpha}(x) =2\, \sum_{l=e,\mu,\tau} \bar \nu_{lL}(x)
\gamma_{\alpha}l_{L}(x) ;~~~ j ^{\mathrm{NC}}_{\alpha}(x)
=\sum_{l=e,\mu,\tau} \bar \nu_{lL}(x) \gamma_{\alpha}\nu_{lL}(x)
\end{equation}
\item
The fields of neutrinos with definite masses enter into CC and NC in
{\em the mixed form}
\begin{equation}\label{4}
\nu_{l L}(x)=\,\sum^{3}_{k=1}U_{l k}\,\nu_{k L}(x).
\end{equation}
Here $\nu_{k }(x)$ is the field of neutrino with mass $m_{k}$ and
$U$ is the unitary PMNS matrix \cite{BP,MNS}.
\end{enumerate}
In the case of neutrino mixing (\ref{4}), the flavor lepton numbers
$L_{e}$, $L_{\mu}$ and $L_{\tau}$ are not conserved. For the
three-neutrino mixing the standard probability of the transition
$\nu_{l} \to \nu_{l'} $ is given by (see \cite{BGG})
\begin{equation}\label{5}
P(\nu_{l} \to\nu_{l'}) = |\sum^{3}_{k=1} U_{l' k} \,e^{-i \Delta
m^2_{ 1k }\frac{L }{2 E }}\, U^{*}_{lk}|^{2},
\end{equation}
where $L$ is the distance between the neutrino-detection and
neutrino-product\-ion points and $E$ is the neutrino energy. Taking into
account the unitarity of the mixing matrix we can rewrite  (\ref{5}) in
the following form
\begin{equation}\label{6}
P(\nu_{l} \to\nu_{l'})=|\delta_{l' l}+ \sum_{k \ne 1} U_{l' k}
\,(e^{-i \Delta  m^2_{  1k }\frac{L }{2 E }} -1) U^{*}_{lk}|^{2}
\end{equation}
The probabilities $P(\nu_{l} \to\nu_{l'})$ in general depend on six
parameters. However, because of the smallness of the parameters
$\frac{\Delta  m^2_{  12 }} {\Delta m^2_{ 23 }}$ and
$\sin^{2}\theta_{13}$ in the leading approximation neutrino
oscillations in the atmospheric-LBL and solar-KamLAND regions are
described by the  simplest two-neutrino expressions which depend,
correspondingly,  on $\Delta  m^2_{ 23 },~ \sin^{2}2\theta_{23}$ and
$\Delta  m^2_{1 2 },~\tan^{2}\theta_{12}$ (see review \cite{BGG}).
The numerical values of these parameters given by (1) and (2) were obtained
from the analysis of the experimental data by using two-neutrino
expressions.

Several derivations of  Eq. (\ref{5}), which
are based on different physical assumptions, exist in the literature. The aim of this paper
is to propose a way of testing these assumptions. We will show
that an experiment using recoilless resonant antineutrino emission and capture, proposed recently
\cite{Raghavan, Potzel}, could provide such a possibility.

\section{Different approaches to neutrino oscillations}
We discuss here different points of view on the physics of
neutrino oscillations.

Neutrinos are produced in CC weak processes. For the difference of
momenta of neutrinos with masses $m_{k}$ and $m_{i}$ (in the
rest-frame of the source) we have
\begin{equation}\label{7}
\Delta p_{ik}= (p_{k}- p_{i}) \sim \frac{\Delta m^{2}_{ik}}{E},
\end{equation}
where $E$ is the neutrino energy (in standard neutrino oscillation
experiments $E\gtrsim$ MeV). From (\ref{7}), (\ref{1}), and (\ref{2})
follows that $|\Delta p_{ik}|$ is much smaller than the
quantum-mechanical uncertainty of the momentum. Thus, it is impossible
to distinguish the emission of neutrinos with different masses in
neutrino-production processes. The matrix element of the production
of $\nu_{k}$ together with lepton $l^{+}$ in a process $a\to b
+l^{+} +\nu_{k}$ ($a$ and $b$ are some hadrons) has the form
\cite{BilG}
\begin{equation}\label{8}
    \langle l^{+}~ \nu_{k}~ b~|S|~a \rangle \simeq
   U^{*}_{lk}~ \langle l^{+} ~\nu_{l}~ b~|S|~a \rangle_{SM},
\end{equation}
where $\langle l^{+} \nu_{l}~ b~|S|~a \rangle_{SM}$ is the Standard
Model matrix element of the process\
\begin{equation}\label{9}
a\to b +l^{+} +\nu_{l}
\end{equation}
in which neutrino masses can be safely neglected.

From (\ref{9}) it follows:
\begin{enumerate}
\item
The state of the flavor neutrino $\nu_{l}$ which is produced in a CC weak
process together with $l^{+}$ is given by
\begin{equation}\label{10}
|\nu_{l}\rangle =\sum_{k}U_{l k}^*~|\nu_{k}\rangle,
\end{equation}
where $|\nu_{k}\rangle$ is the state of a neutrino with mass $m_{k}$
and 4-momentum $p_{k}=(E_{k}, \vec{p_{k}})$.

\item
The probabilities of the processes of neutrino production are given
by the SM.
\end{enumerate}

Thus, in a charged-current weak process a flavor neutrino $\nu_{l}$, 
which is described by the state (\ref{10}), is produced. What will
be the state of the neutrino after some time $t$ (at some distance $L$)?
Two different approaches to the propagation of neutrino states are discussed in the literature.

\begin{center}
 {\bf I. Evolution in time.}
\end{center}
The evolution equation of any quantum system is the Schr\"odinger equation
(see, for example, \cite{BogShir})
\begin{equation}
 i\,~ \frac{\partial |\Psi(t)\rangle}{\partial t} =
H\,~ |\Psi(t)\rangle     \,. \label{11}
\end{equation}
Here $|\Psi(t)\rangle$ is the state of the system at the time $t$
and $H$ is the total Hamiltonian. The general solution of this
equation has the form
\begin{equation}\label{12}
|\Psi(t)\rangle = e^{-i\,H t}\,|\Psi(0)\rangle,
\end{equation}
where $|\Psi(0)\rangle $ is the state of the system at the initial
time $t=0$.

If $|\Psi(0)\rangle =|\nu_{l}\rangle $, we have for the neutrino state in
vacuum at the time $t$
\begin{equation}\label{13}
|\nu_{l}\rangle_{t}=e^{-i\,Ht}\,|\nu_{l}\rangle=
\sum_{k}e^{-iE_{k}t}\,~U_{lk }^*|\nu_{k}\rangle.
\end{equation}
Thus, if the energies $E_{k}$ are different, the neutrino state
$|\nu_{l}\rangle_{t}$ is a \textit{non-stationary} one. For such states
the time-energy uncertainty relation
\begin{equation}\label{14}
\Delta E~ \Delta t \gtrsim 1
\end{equation}
holds (see, for example, \cite{Sakurai}). In this relation
$\Delta E$ is the energy uncertainty and $\Delta t $ is the time
interval during which the state of the system is significantly
changed.

Neutrinos are detected via the observation of CC and NC reactions. In
such reactions, flavor neutrinos $\nu_{l'}$, which are described by
mixed coherent states (\ref{10}), are detected. From (\ref{10}) and
(\ref{13}) we find
\begin{equation}\label{15}
|\nu_{l}\rangle_{t}=\sum_{l'}|\nu_{l'}\rangle~\sum_{k}
U_{l'k}~e^{-iE_{k}t}\,~U_{lk }^*.
\end{equation}
Thus, the transition probability $\nu_{l}\to \nu_{l'}$ is given by
\begin{equation}\label{16}
P(\nu_{l}\to \nu_{l'}) =|\sum^{3}_{k=1}
U_{l'k}~e^{-i(E_{k}-E_{1})t}\,~U_{lk }^*|^{2}.
\end{equation}
From this expression it is obvious that in the case of equal
energies of the neutrinos with different masses $P(\nu_{l}\to
\nu_{l'})=\delta_{l'l}$. {\em Thus, in the approach based on the
Schr\"odinger evolution equation, there will be no neutrino oscillations
if $E_{k}=E_{i}$} \cite{BilenkyMat}.

Let us assume now that the flavor neutrino states $ |\nu_{l}\rangle $
are superpositions of the neutrino states $\nu_{k}$ {\em with
the same momentum $\vec{p}$.}  In this case we have
\begin{equation}\label{17}
E_{k} =\sqrt{p^{2}+m^{2}_{k}}\simeq p+\frac{m^{2}_{k}}{2E},
\end{equation}
with $E$ being the neutrino energy at $m_{k}\to 0$.

Taking into account that for ultrarelativistic neutrinos
\begin{equation}\label{18}
t\simeq L
\end{equation}
we obtain from (\ref{16}) the standard expression (\ref{5}) for the
transition probability which perfectly describe existing neutrino
oscillation data. The time-energy uncertainty relation (\ref{14}) takes the form of the well-known condition for the observation of neutrino oscillations:
\begin{equation}\label{19}
(E_{k}-E_{1})t \simeq \frac{\Delta m^{2}_{1k}}{2E}L\gtrsim 1.
\end{equation}

Let us stress again that in the approach based on the Schr\"odinger
evolution equation, oscillations between different flavor neutrinos
are due to the fact that the neutrino state $|\nu_{l}\rangle_{t}$ is a
superposition of states with \textit{different} energies \cite{BilenkyMat}. \footnote{ We
assumed that the states of flavor neutrinos are superpositions of states
of neutrinos with different masses and the same momentum. Let us
notice that if we assume that neutrinos with different masses have
different momenta, in the expression for the transition probability we
will have terms $(p_{k}- p_{1})L$ in addition to the standard phases $\frac{\Delta m^{2}_{1k}}{2E}L$. These additional terms could be of the same order
as the standard phases and could be different in different
experiments. All analyses of neutrino oscillation data  do not favor
such a possibility.}

\newpage
\begin{center}
 {\bf II. Evolution in time and space.}
\end{center}

It has been suggested in several papers (see
\cite{Lipkin,Kayser,Giunti}) that the mixed neutrino state at the
space-time point $x=(t, \vec{x})$ is given by
\begin{equation}\label{20}
|\nu_{l}\rangle_{x}= \sum^{3}_{k=1}e^{-ip_{k}x}\,~U_{lk
}^*|\nu_{k}\rangle.
\end{equation}
Here  $|\nu_{k}\rangle$ is the state of a neutrino with mass $m_{k}$
and momentum $p_{k}$. From  (\ref{20}) we find
\begin{equation}\label{21}
|\nu_{l}\rangle_{x}=e^{-ip_{1}x}
\sum_{l'}|\nu_{l'}\rangle~\sum^{3}_{k=1}
U_{l'k}~e^{-i(p_{k}-p_{1})x}\,~U_{lk }^*.
\end{equation}
Thus,
\begin{equation}\label{22}
P(\nu_{l}\to \nu_{l'})=|\sum_{k}
U_{l'k}~e^{-i(p_{k}-p_{1})x}\,~U_{lk }^*|^{2}
\end{equation}
is the probability to find the flavor neutrino $\nu_{l'}$ at the point
$x$ in the case that at point $x=0$ the mixed flavor neutrino
$\nu_{l}$ was produced. For the phase difference we have
\begin{equation}\label{23}
(p_{k}-p_{1})x=(E_{k}-E_{1})~t-(p_{k}-p_{1})~L=
\frac{E^{2}_{k}-E^{2}_{1}}{E_{k}+E_{1}}t- (p_{k}-p_{1})L,
\end{equation}
where $\vec{p_{k}}=p_{k}\vec{k}$ and $\vec{k}\vec{x}=L$ with
$\vec{k}$ being the unit vector in the direction of the momenta.

In the framework of the evolution of the flavor states in time and
space two scenarios were considered.
\begin{center}
\textbf{Scenario I.}~~ $t\simeq L.$
\end{center}
From (\ref{23}) we find for the oscillation phase
\begin{equation}\label{24}
(p_{k}-p_{1})x=\frac{E^{2}_{k}-E^{2}_{1}}{E_{k}+E_{1}}t-
(p_{k}-p_{1})L \simeq  (p_{k}-p_{1})(\frac{p_{k}+p_{1}}{E_{k}+E_{1}}t
-L)+\frac{\Delta m^{2}_{1k}}{2E}t
\end{equation}
If we assume now that the distance and the time are connected by the
relation (\ref{18}), we find from (\ref{24}) (up to terms linear in
$m^{2}_{k}$) the standard expression (\ref{5}) for the
transition probability with the oscillation phase
\begin{equation}\label{25}
(p_{k}-p_{1})x\simeq \frac{\Delta m^{2}_{1k}}{2E}L.
\end{equation}
This result is valid if $p_{i}\neq p_{k}$. The last
inequality means that 1. $E_{k}\neq E_{i};~~\vec{p_{k}}=\vec{p_{i}}$
or 2. $E_{k}= E_{i};~~\vec{p_{k}}\neq\vec{p_{i}}$ or 3. $E_{k}\neq
E_{i};~~\vec{p_{k}}\neq \vec{p_{i}}$.

\begin{center}
\textbf{Scenario II.}~~Stationary states.
\end{center}
It has been suggested in \cite{Stodol,Lipkin,Kayser} that time is not
measured in neutrino oscillations and that neutrinos with different
masses have the same energies, i.e. the neutrino state is stationary. 
Taking into account that in this case the momenta of the neutrinos $\nu_{k}$ and
$\nu_{i}$ are different we will come to the standard
expression (\ref{5}) for the transition probability between
different flavor neutrinos with the oscillation phase
\begin{equation}\label{26}
(p_{k}-p_{1})x=\frac{\Delta m^{2}_{1k}}{2E}L
\end{equation}
Thus, in all three cases which we have considered we came to the same
expression (\ref{5}) for the transition probability. This means that in
usual neutrino oscillation experiments it is impossible to
distinguish these three cases.

Recently, a new type of neutrino experiment based on the M\"ossbauer effect has been proposed
\cite{Raghavan,Potzel}. In the next section we
will discuss this proposal from the point of view that it might provide a possibility
to distinguish the different assumptions on the evolution of
mixed neutrino states.

\section{Recoilless resonant emission and absorption of antineutrinos}

In \cite{Raghavan}, an experiment has been proposed for the detection of
$\bar\nu_{e}$ with energy $\simeq$ 18.6 keV in recoilless resonant (M\"ossbauer)
transitions:
\begin{equation}\label{27}
 ^{3}\rm{H}\to ^{3}\rm{He}+\bar\nu_{e};~~~\bar\nu_{e}+
^{3}\rm{He}\to ^{3}\rm{H}.
\end{equation}
For the cross section of recoilless resonant absorption of $\bar\nu_{e}$
by $^{3}\rm{He}$ the value $\sigma_{R} \simeq
3\cdot 10^{-33}\rm{cm}^{2}$ was obtained \cite{Raghavan}. With the aim to
determine the value of the parameter $\sin^{2}\theta_{13}$ it
was proposed to study neutrino oscillations, driven by $\Delta
m^{2}_{23}$, in an experiment with a baseline of $\sim$ 10 m.

It was estimated in \cite{Raghavan} that the uncertainty of the energy
of antineutrinos is of the order
\begin{equation}\label{28}
\Delta E \simeq 8.6 \cdot 10^{-12}\rm{eV}.
\end{equation}
Let us stress that $\Delta E $ is much smaller than the difference of the
energies between $\nu_{3}$ and $\nu_{2}$
\begin{equation}\label{29}
(E_{3}-E_{2})\sim \frac{\Delta m^{2}_{23}}{2E}\simeq 6.5 \cdot
10^{-8}\rm{eV},
\end{equation}
which drive neutrino oscillations in the case of the approach based
on evolution in time.

The state of flavor $\bar\nu_{e}$ produced and detected in the reactions (\ref{27})
is the superposition of states of neutrinos with the same energy
and different momenta. Thus, in the experiment proposed in
\cite{Raghavan} neutrino oscillations will not be observed if
the approach based on the Schr\"odinger evolution equation is correct.
On the other hand, neutrino oscillations can be observed
in this experiment if one of the Scenarios I or II, based on the evolution in
space and time, is correct. Thus, an experiment with recoilless resonant emission and 
absorption of antineutrinos could have an important impact on our
understanding of the physics of neutrino oscillations.

In conclusion we make the following remarks:

\begin{enumerate}

\item

In accelerator experiments K2K \cite{K2K} and MINOS \cite{Minos} the
time of neutrino production and neutrino detection is measured. In
the K2K experiment neutrino events which satisfy the criteria
\begin{equation}\label{30}
  -0.2\leq |\Delta t-\frac{L}{c}|\leq 1.3~\mu s
\end{equation}
were chosen. Here $\Delta t=t_{SK}-t_{KEK}$ ($t_{SK}$ is the time
of detection of neutrino events in the Super-Kamiokande detector and
$t_{KEK}$ is the time of the production of neutrinos at KEK). Thus,
neutrino oscillations are a phenomenon with a finite time difference
between neutrino detection and neutrino production. According to the
time-energy uncertainty relation, in neutrino oscillations $\Delta
E$ must be different from zero, i.e. neutrino oscillations are a {\em
non-stationary} process. These arguments are in favor of the
approach based on evolution in time (see Section 2I) and scenario I (see Section 2II).
\item

In a disappearance experiment, as described in \cite{Raghavan}, with recoilless resonance absorption of 18.6 keV antineutrinos by $^{3}\rm{He}$ a positive effect of neutrino
oscillations can be observed only in the case that the parameter
$\sin^{2}\theta_{13}$ is not too small. At present, only an upper bound
$\sin^{2}\theta_{13}\leq 5\cdot 10^{-2}$ is known from the data of
the CHOOZ experiment\cite{Chooz}. A positive result of an
oscillation experiment with recoilless resonant antineutrino absorption would
allow to determine the parameter $\sin^{2}\theta_{13}$ and to
exclude the approach based on the evolution of the mixed neutrino states
in time (see Section 2I). However, a negative result of such an experiment could be
the consequence of the smallness of the parameter $\sin^{2}\theta_{13}$.
Thus, in the case of a negative result of an experiment with recoilless resonant
antineutrino absorption, definite conclusions on the fundamentals of
neutrino oscillations can be drawn only if the parameter
$\sin^{2}\theta_{13}$ will be measured in future reactor (DOUBLE
CHOOZ \cite{2Chooz}, Daya Bay \cite{Dayabay}) or accelerator (T2K
\cite{T2K}, Nova \cite{Nova}) experiments.

\item
We have considered an experiment, proposed in \cite{Raghavan}, on the search for neutrino oscillations driven by the 'large' $\Delta m^{2}_{23}$ and based on recoilless resonant absorption of $\bar \nu_{e}$. The baseline of this experiment is $\thicksim$ 10 m. The effect of neutrino
oscillations will be small (if present at all) because the amplitude of the oscillations is limited  by the upper bound of the CHOOZ experiment ($\sin^{2}2\theta_{13}\lesssim 2\cdot 10^{-1}$). The question arises if a similar oscillation experiment could be performed which, however, is driven by the 'small' $\Delta m^{2}_{12}$. Such an experiment would require an about 30 times larger baseline, i.e., $\thicksim$ 300 m. Because tritium acts as a point-like source the expected number of neutrino events in such an experiment will be $\thicksim$ 1000 times smaller than in the 10 m-baseline experiment.

It was estimated in \cite{Raghavan} that with a 10 m baseline about $10^{3}$ $\bar\nu_{e}$-captures/day can be expected. Thus, in an experiment with a baseline of $\thicksim$ 300 m only about 1 capture/day could be observed. If we neglect the small contributions of the terms proportional
to $\sin^{2}\theta_{13}$ we find for the $\bar\nu_{e}$ survival probability
from eq.(\ref{5})
\begin{equation}\label{31}
P(\bar\nu_{e}\to \bar\nu_{e}) =1 -\frac{1}{2}\sin^{2}2\theta_{12}
(1-\cos\Delta m^{2}_{12}\frac{L}{2E}).
\end{equation}
Taking into account that the amplitude of the neutrino oscillations is large in this case (see eq.(\ref{2})), such an experiment might still be feasible although we regard the estimate given in \cite{Raghavan} as rather optimistic \cite{Potzel}.

The uncertainty of the energy of the antineutrinos emitted without recoil, given by
eq.(\ref{28}), is much smaller than
\begin{equation}\label{32}
\frac{\Delta m^{2}_{12}}{2E}\simeq 2.1\cdot 10^{-9}\rm{eV}.
\end{equation}
Thus, all our arguments given above for the possibilities to distinguish different
approaches to the physics of neutrino oscillations are applicable also in
this case. Notice that two detectors of the same kind would allow to record the antineutrinos at two distances ($\thicksim$ 10 m and $\lesssim$ 300 m).

\item
Effects of  small neutrino masses can not be revealed in neutrino
production and neutrino detection experiments with neutrino and
antineutrino energies $\gtrsim $ 1 MeV. Hence, the states of flavor
neutrinos $\nu_{e}, \nu_{\mu}, \nu_{\tau}$ and flavor antineutrinos
$\bar\nu_{e}, \bar\nu_{\mu}, \bar\nu_{\tau}$ which are produced and
detected in corresponding processes, are {\em mixed states}.
Neutrino oscillations take place because {\em in the propagation of
states of neutrinos with definite masses} small neutrino
mass-squared differences are relevant and can be determined if $L/E$
is large enough. We considered here different assumptions on the propagation of
neutrino states (in time or in space and time). Different
assumptions on the propagation of neutrino states give the same
transition probabilities in the case of standard neutrino
oscillation experiments. We have shown that a recently proposed
M\"ossbauer-type neutrino experiment \cite{Raghavan,Potzel} could allow to
distinguish the different fundamental assumptions on the propagation of
neutrinos with definite masses.

\end{enumerate}

We thank the anonymous referee for suggesting to consider also the neutrino mass-squared difference associated with solar neutrino oscillations. We thank T. Schwetz for fruitful discussions. S. Bilenky acknowledges the ILIAS program for support.

\end{document}